\documentclass[12pt]{iopart}
\usepackage[english]{babel}
\usepackage{graphicx}
\begin{document}
\title{Stochastic background from extra-galactic double neutron stars}
\author{T. Regimbau and B. Chauvineau}
\address{Dpt. ARTEMIS \\Observatoire de la C\^ote d'Azur, \\ BP 429
06304 Nice, France}
\ead{regimbau@oca.eu;chauvineau@oca.eu}

\begin{abstract}

We present Monte Carlo simulations of the extra galactic population of
inspiralling double neutron stars, and estimate its contribution to
the astrophysical gravitational wave background, in the frequency range
of ground based interferometers, corresponding to the last thousand seconds
before the last stable orbit when more than 96 percent of the signal
is released. We show that sources at redshift $z>0.5$
contribute to a truly continuous background which may be detected by
correlating third generation interferometers.

\end{abstract}

\section{Introduction}

Double neutron stars (DNSs) are very promising sources of
gravitational waves for both ground based interferometers, which are
sensitive to the last phase of the coalescence and the collapse, and
the space detector LISA, which is expected to detect the continuous
low frequency inspiral phase.
In a previous work, we presented new estimates of the
merging rate in the local Universe (including the contribution from
ellipticals and taking into account the star formation history derived
directly from observation) and discussed its consequences for the
first generations of ground based interferometers. We predicted a
detection every 148 and 125 years with Virgo and LIGO in their
initial configuration, and up to 6 detections per year in their advanced configuration \cite{reg05,dfp06}.
In addition to the emission from the nearest DNSs, 
it is expected that the superposition of a large number of unresolved
sources produces a stochastic background.
The background from the low frequency inspiral phase of various
populations of compact binaries, which represents the main source of
confusion noise for LISA, was studied intensively in the past decades
(see for instance
\cite{eva87,hil90,ben97,pos98,nel01} for the galactic foregrounds, and
\cite{ign01,sch01,far02,coo04} for the extra-galactic contribution).
These predictions usually rely on binary evolution codes to estimate initial
parameters such as eccentricity, mass ratio, orbital separation, which
introduce large uncertainties due to the difficulty modelling mass
loss and mass exchanges.
In this work, we investigate the stochastic background produced by the
extra-galactic population of DNSs,
carrying on with the the previous estimates of
\cite{reg06,reg07}. We are interested in the last phase of the
coalescence, up to the last stable orbit (LSO), at kHz frequencies,
when the system is in circular orbit and when most of the GW energy is
released.
The article will be organized as follow:
In section 1, we present a direct calculation of the 
spectral properties of the stochastic background; in section 2, we
discuss the detection of the background with different generations of
terrestrial interferometers; in section 3, we present Monte Carlo
simulations of the extra-galactic population of DNSs and its
contribution to the GW background; 
and finally in section 5, we summarize our results and discuss
the interest of the Monte Carlo simulations for
source modelling and data analysis, as well as possible improvements.

\section{The GW background}

The spectrum of the gravitational stochastic background is
characterized by the dimensionless density parameter (or closure
density) \cite{all97a,mag00}:

\begin{equation}
\Omega_{gw}(\nu_o)=\frac{1}{\rho_c}\frac{d\rho_{gw}}{d\ln \nu_o}
\label{eq-omega_general}
\end{equation}

where $\rho_{gw}$ is the gravitational energy density, $\nu_o$ the
frequency in the observer frame and $\rho_c=\frac{3H_0^2}{8 \pi G}$
the critical energy density needed to close the Universe today.
Throughout this paper, we assume a flat Einstein de Sitter 737
cosmology, with energy density of matter$\Omega_m=0.3$, energy density
of vacuum $\Omega_{\Lambda}=0.7$ and Hubble
parameter $H_0=70$ km s$^{-1}$ Mpc$^{-1}$ \cite{rao06}, corresponding
to the so-called concordant model derived from observations of distant
type Ia supernovae \cite{per99}  and the power spectra of the cosmic
microwave background fluctuations \cite{spe03}.

For a stochastic background of astrophysical origin  \cite{fer99a}:

\begin{equation}
\Omega_{gw}=\frac{1}{\rho_c c^3} \nu_o F_{\nu_o}
\label{eq-omega_astro}
\end{equation}

where the integrated flux (in erg cm$^2$ Hz$^{-1}$ s$^{-1}$) at the observed frequency $\nu_o$ is defined
as:

\begin{equation}
F_{\nu_o}=\int f_{\nu_o} \frac{dR^o}{dz} dz
\label{eq-flux}
\end{equation}

The spectral properties of a single source located at redshift $z$ are given by
the fluence (in erg cm$^2$ Hz$^{-1}$) \cite{pea99,fer99a}: 

\begin{equation}
f_{\nu_o}=\frac{1}{4 \pi d_L^2} \frac{dE_{gw}}{d \nu_o}=\frac{1}{4 \pi
  d_L^2} \frac{dE_{gw}}{d \nu}(1+z)
\label{eq-fluence}
\end{equation}

where $d_L=r(1+z)$ is the luminosity distance, $r$ the proper
distance, which depends on the adopted cosmological model,
and $\nu=\nu_o(1+z) $ the frequency in the source frame.
In the quadrupolar approximation, the spectral GW energy emitted by a
binary system, which inspirals in a circular orbit is given by \cite{pet63,mis95}: 
\begin{equation}
dE_{gw}/{d\nu} = K \nu^{-1/3} 
\label{eq-energy}
\end{equation} 
where
\begin{equation}
K = \frac{(G \pi)^{2/3}}{3} \frac{m_1m_2}{(m_1+m_2)^{1/3}}
\label{eq-K}
\end{equation}
For double neutron stars with masses $m_1 = m_2 =
1.4$ M$_\odot$, one obtains $K=5.2 \times 10^{50}$ erg
Hz$^{-2/3}$ and the gravitational frequency at the last stable orbit is
assumed to be $\nu_{LSO}=1.5$ kHz \cite{sat02}.

The merging of two neutron stars (NSs) occurs long after the formation of the
progenitors and this delay must be taken into account when calculating the event rate.
The progenitor formation rate per comoving volume is given on the
time scale of the observer by:
\begin{equation}
R^o_{f}(z_f)= \lambda \frac{R_*(z_f)}{1+z_f}
\label{eq-vrate_prog}
\end{equation}
where $R_*(z_f)$ is the cosmic star formation rate per comoving volume
(SFR) at the time of formation $z_f$ and is expressed in M$_\odot$ Mpc$^{-3}$ yr$^{-1}$. 
In our calculations, we consider the recent model
of \cite{hop06}, constrained by the Super Kamiokande limit on the electron
antineutrino flux from past core-collapse supernovas up to $z_{\max}=6$. 
The $(1+z)$ term in the denominator corrects the cosmic star formation rate by the time dilatation due to the cosmic expansion. 
The factor $\lambda$ is the mass fraction
converted into the progenitors, assumed to be the same at all redshifts,
and given by \cite{dfp06} as the product
$\lambda=\beta_{NS}f_b\lambda_{NS}$, where $\beta_{NS}$ is the
fraction of binaries which remains bounded after the second supernova
event, $f_b$ the fraction of massive binaries formed among all stars
and $\lambda_{NS}$ the mass fraction of NS progenitors derived from a
modified Salpeter A IMF with minimal and maximal initial masses of 8 M$_{\odot}$ and 40 M$_{\odot}$, as suggested by \cite{hop06}.
Numerically one obtains $\lambda=3 \times 10^{-5}$ M$_\odot^{-1}$. 

The coalescence rate per comoving volume on the time scale of the observer results from the convolution between $R^o_f$ and the
probability distribution of the coalescence time $\tau_c$ (or the delay
between the formation of the DNS after the second supernova explosion
and the coalescence), which depends on
the orbital parameters (separation and eccentricity) and
neutron star masses, at the time the DNS is formed, and is given by 
\cite{dfp06} as:
\begin{equation}
P_{\tau_c}(\tau_c)=\frac{0.087}{\tau_c} \,\ {\rm  with
  }\,\ \tau_c \in [0.2 \rm{Myr} - 20 \rm{Gyr}]
\label{eq-proba_tau}
\end{equation}
 
In terms of the cosmic (lookback) time:
\begin{equation}
T_c = T (z_c)= \int^{z_c}_0 \frac{1}{H_0} \frac{dz}{E(z)(1+z)} 
\label{eq-cosmictime}
\end{equation}
where 
\begin{equation}
E(\Omega,z)=\sqrt{\Omega_{\Lambda}+\Omega_{m}(1+z)^3}
\end{equation}
and where $z_c$ is the redshift of coalescence,
it writes:
\begin{equation}
R^o_c(z_c) = R^o_c(T_c) = \int R^o_f(T_c +\tau_b + \tau_c)
P_{\tau_c}(\tau_c) d\tau_c
\label{eq-vrate_coal}
\end{equation}
where $\tau_b$ ($\sim 10^8$ yr) is the mean lifetime of the
progenitors (or the time for the two massive stars to evolve into two
neutron stars) \cite{dfp06}.
The lower limit of the integral corresponds to the minimal coalescence
time $\tau_o=0.2$ Myr \cite{dfp06}, while the upper limit is fixed by the maximal redshift of the
adopted star formation rate and is given by the
maximal formation time delayed according to the time to reach the
coalescence, namely $T(z_{\max})-\tau_b - T_c$.
For $z_{\max}=6$, $T(z_{\max})=T_{\max}=12.5$ Gy.  
\begin{figure}
\centering
\includegraphics[angle=0,width=0.8\columnwidth]{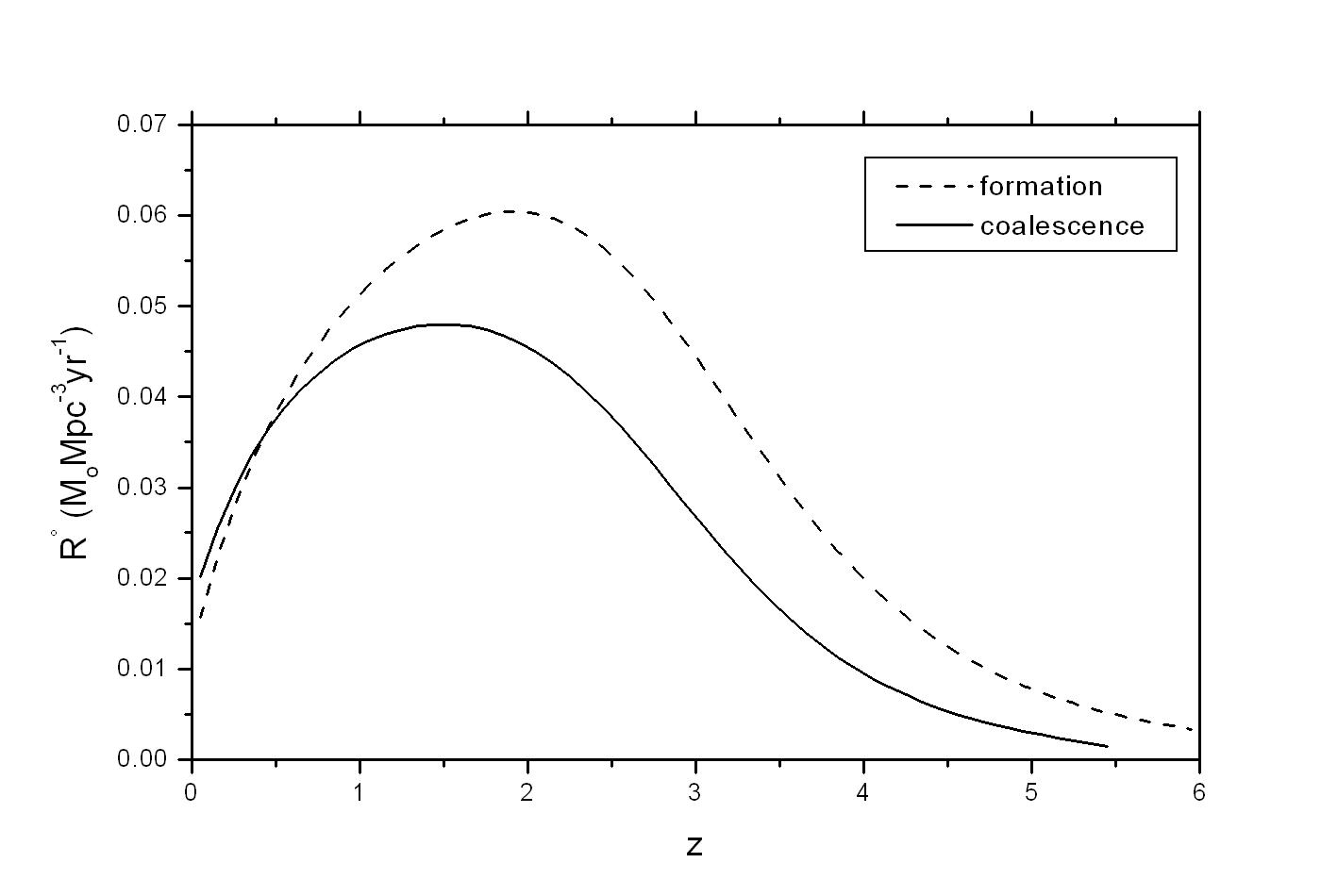}
\caption{comparison between the formation rate of DNS progenitors and
  the coalescence rate.
\label{fig-DNS_Rzc}} 
\end{figure}

As one can see in Fig.~\ref{fig-DNS_Rzc}, the
coalescence rate is shifted toward lower redshifts, with respect to
the formation rate, reflecting the time delay between the formation of
the progenitors and the coalescence event.

The event rate per redshift interval in eq.~\ref{eq-flux}, is thus
given by multiplying $R^o_c(z)$ by the element of comoving volume:
\begin{equation}
\frac{dR^o}{dz}= R^o_c \frac{dV}{dz}
\label{eq-rate}
\end{equation} 
where
\begin{equation}
\frac{dV}{dz}=4 \pi r^2 \frac{c}{H_0} \frac{1}{E(\Omega,z)}
\end{equation}

Combining the expressions above and replacing the constants by their usual values, one obtains:

\begin{equation}
\Omega_{gw}= 8.6 \times 10^{-10} \nu_o^{2/3}
\int^{z_{\sup}}_0 \frac{R^o_c(z)}{(1+z)^{4/3}E(z)} dz
\label{eq-omega_DNS}
\end{equation} 

The upper limit of the integral, which depends on both the maximal
emission frequency in the source frame $\nu_max$ and the maximal redshift of the
model of star formation history ($z_{\max} \sim 6$), is given by: 

\begin{equation}
z_{\sup}= 
\left\lbrace
\begin{array}{ll} 
z_{\max}    &   \hbox{  if } \nu_o < \frac{\nu_{\max} }{1+z_{\max}}\\
\frac{\nu_{\max}}{\nu_o}-1 &   \hbox{  otherwise }\\
\end{array} 
\right.
\label{eq-zsup}
\end{equation}

Consequently, the shape of the spectrum is
characterized by a cutoff at the maximal emission frequency and a
maximum at a frequency which depends on the shape of both the SFR and
the spectral energy density. Before the maximum,
$\Omega_{gw}$ increases as $\nu_o^{2/3}$ (eq.~\ref{eq-omega_DNS}).

Besides the spectral properties, it is important to study the nature of
the background \cite{cow06}. In the case of burst sources the
integrated signal received at $z=0$ from sources up to redshift $z$,
would show very different statistical behaviour whether the duty cycle
\cite{mag00}: 

\begin{equation}
D(z)=\int^z_0 \bar{\tau} (1+z') \frac{dR^o}{dz}(z') dz'
\label{eq-DC}
\end{equation}

defined as the ratio, in the observer frame, of the typical duration
of a single event $\bar{\tau}$, to the average time interval between successive events, is smaller or larger than unity. 
When the number of sources is large enough for the time interval
between events to be small compared to the duration of a single event
($D>> 1$), the waveforms overlap to produce a continuous
background. Due to the central limit theorem, such backgrounds obey
the Gaussian statistic and are completely determined by their spectral
properties. They could be detected by data analysis
methods in the frequency domain such as the cross correlation
statistic presented in the next section \cite{all99}. 
On the other hand, when the number of
sources is small enough for the time interval between events to be
long compared to the duration of a single event ($D << 1$), the
sources are resolved and may be detected by data
analysis techniques in the time domain (or the time frequency domain)
such as match filtering \cite{arn99,pra01}.
An interesting intermediate case arises when the time
interval between events is of the same order of the duration of a
single event. These signals, which sound like crackling popcorn, are
known as "popcorn noise". The waveforms may overlap but the statistic
is not Gaussian  anymore so that the amplitude on the detector at a
given time is unpredictable. For such signals, data analysis
strategies remain to be investigated \cite{dra03}, since the time
dependence is important and data analysis techniques in the frequency
domain, such as the cross correlation statistic, are not adapted. The
critical redshifts at which the background becomes continuous, popcorn
or shot noise will be fixed by the conditions $D(z_c)=$1, 0.1 or 0.01
\cite{cow06}. 

\begin{figure}
\centering
\includegraphics[angle=0,width=0.8\columnwidth]{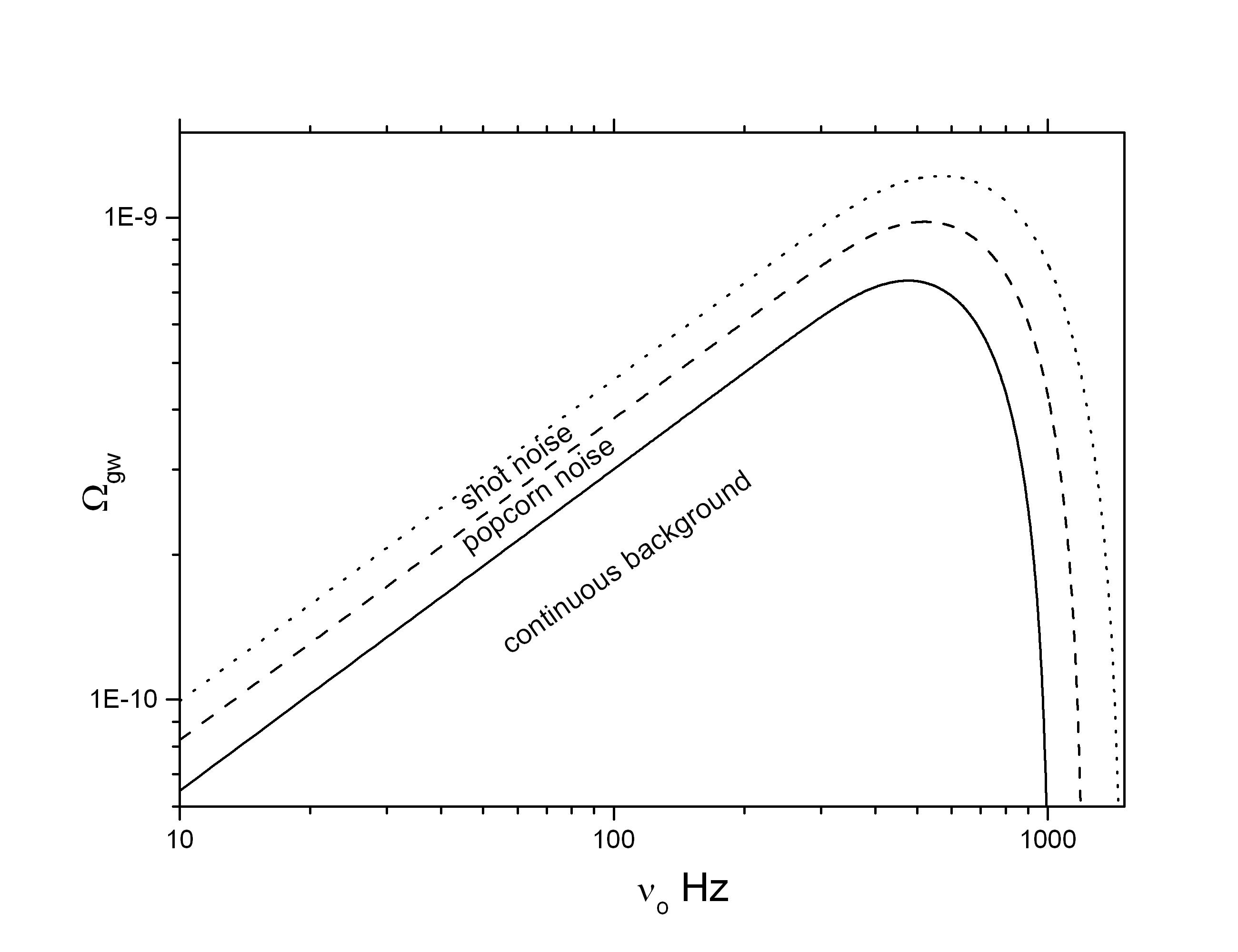}
\caption{closure density of the continuous background produced by DNS
  coalescences at $z > 0.5$ (continuous line) and of the popcorn
  contribution corresponding to sources between $z=0.2-0.5$ (dashed
  line). The signal from the whole population down to $z=0$ is also plotted for
  comparison (dotted line).
\label{fig-DNS_omega}} 
\end{figure}

In our calculations, we considered the last $ \sim
1000$ s before the last stable orbit, when more than 96\% of the
gravitational energy is released and when the signal range between
$10-1500$ Hz, in the frequency domain of ground based detectors
\cite{reg06}.
At that time, the system has been circularized through GW
emission (eq.~\ref{eq-energy}) and all the emission is assumed to take place at the redshift
of coalescence.
\begin{figure}
\centering
\includegraphics[angle=0,width=0.8\columnwidth]{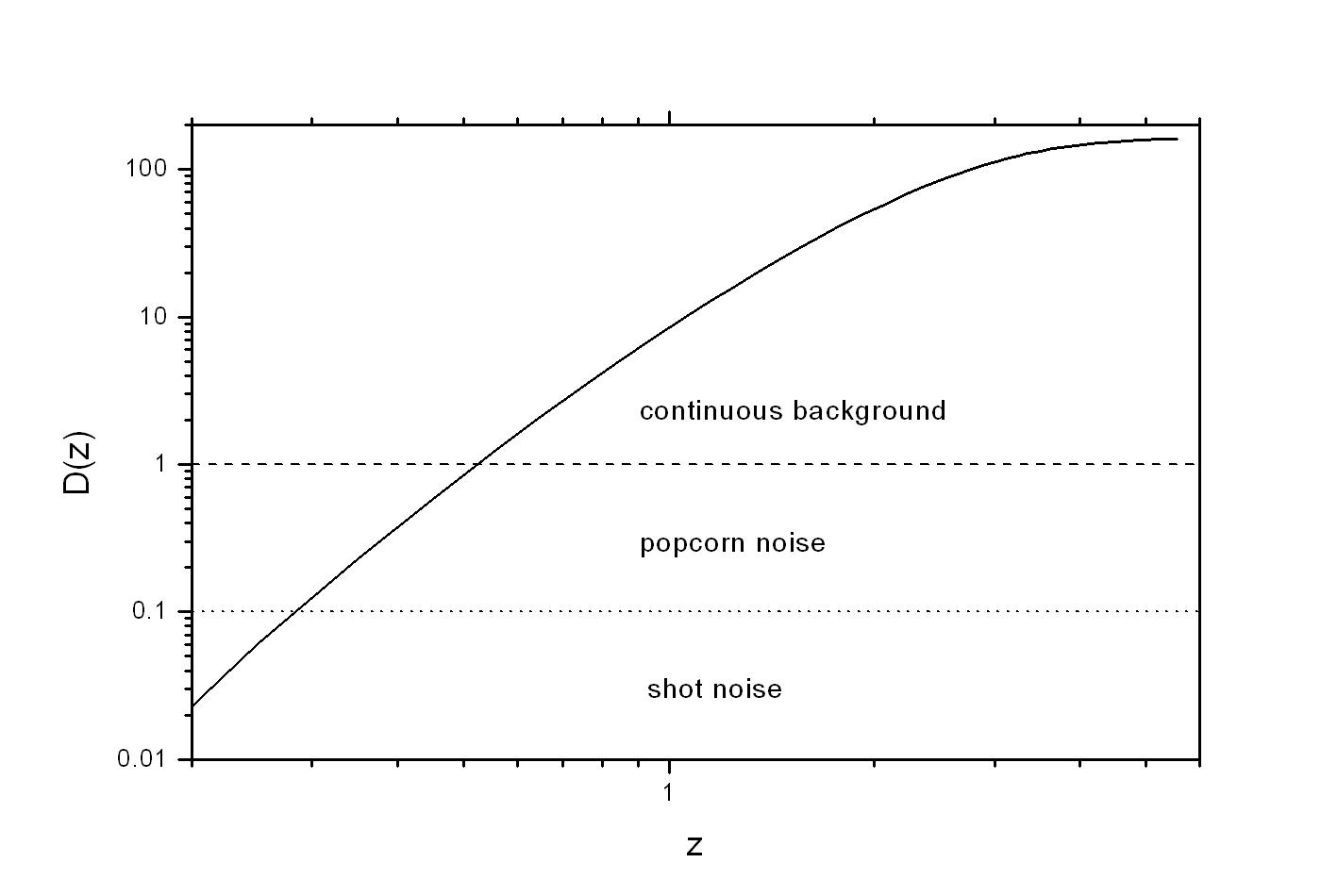}
\caption{duty cycle from sources up to redshift $z$. The horizontal lines
  at $D=1$ and $D=0.1$ show the limit of the continuous and the popcorn backgrounds.  
\label{fig-DNS_DC}} 
\end{figure} 
We show that sources at redshifts $z>0.5$
contribute to a truly continuous stochastic background, while sources at
redshifts $0.2<z<0.5$  are responsible for a popcorn noise, with duty
cycle of 1 and 0.1 respectively (Fig.~\ref{fig-DNS_DC}). 
The closure density reaches a
maximum of $\Omega_{gw} \sim 7.3 \times  10^{-10}$  around  480 Hz for
the continuous contribution and of $\Omega_{gw} \sim 9.7 \times
10^{-10}$ around 520 Hz for the popcorn background (Fig.~\ref{fig-DNS_omega}).
The total background, including the nearest sources down to $z \sim 0$
is slightly higher, with a maximum of $\Omega_{gw} \sim 1.2 \times
10^{-9}$ at 560 Hz.  
\cite{sch01} used a similar procedure to calculate the background in
the frequency range of LISA. At lower frequencies our results are
comparable (with $\Omega_{gw} \sim 4-5 \times 10^{-10}$ at around 100
Hz for the total background) besides different assumptions
about the SFR, the mass range for NS progenitors and the distribution of
the coalescence time. But in \cite{sch01} the maximum occurs at
lower frequencies ($\sim 100$ Hz), since being interested in the
signal in the range between 10 $\mu$Hz and 1 Hz, they have set the
value of the maximum frequency to that expected at a separation three
times the last stable orbit ($\nu_{\max} = 0.19 \nu_{LSO}$). 
Those authors have stressed that their calculations are expected to be
accurate in the frequency band of LISA, thus a direct comparison with
our predictions for the frequency band of ground based detectors is
probably not very meaningful.

\section{Detection}
 
\begin{figure}
\centering
\includegraphics[angle=0,width=0.8\columnwidth]{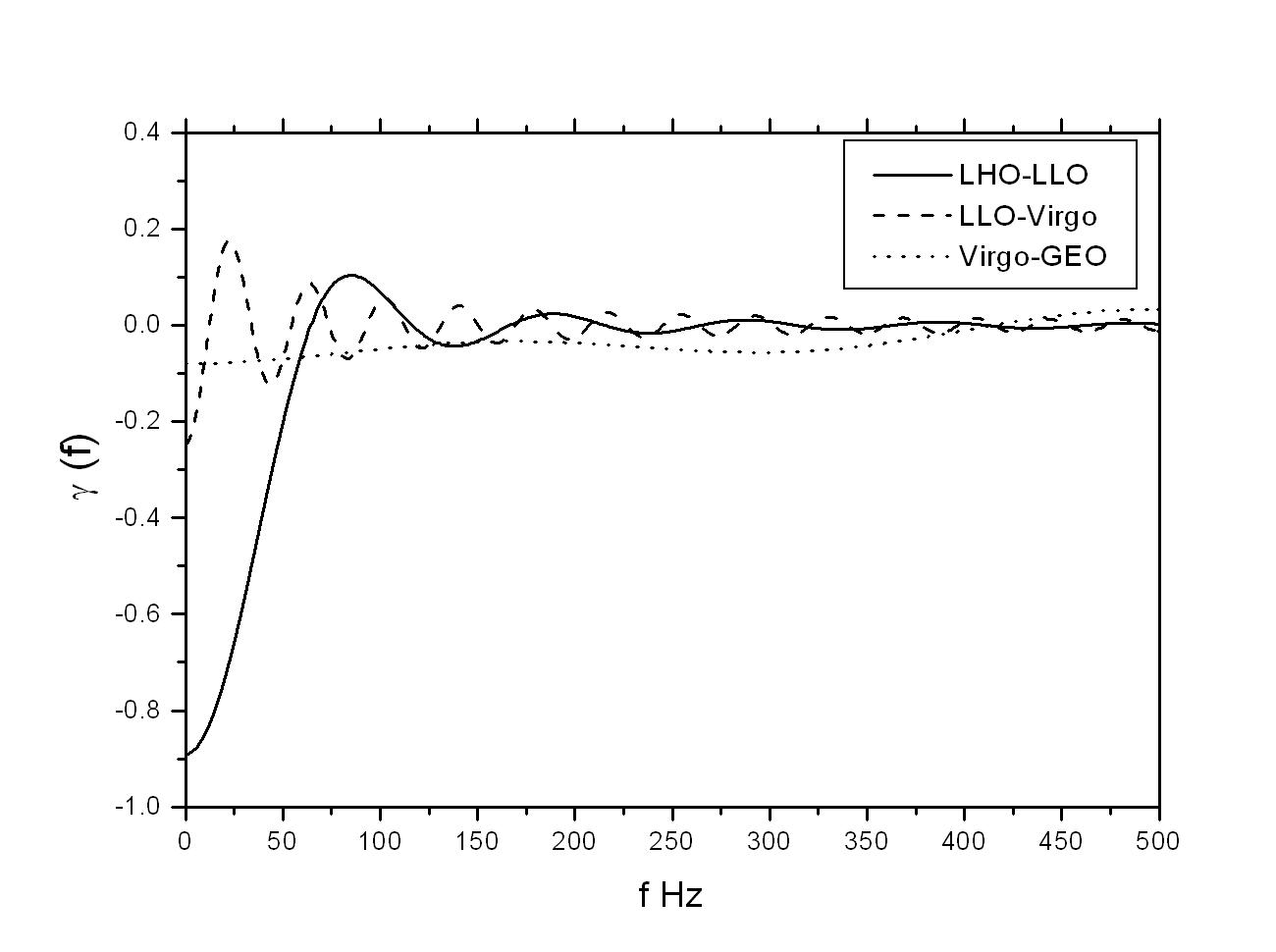}
\caption{overlap reduction function for the most promising detector
  pairs. LHO and LLO stand for LIGO Hanford Observatory and
  LIGO Livingston Observatory.
\label{fig-overlap}} 
\end{figure}

The optimal strategy to search for a gaussian (or continuous) stochastic
background, which can be confounded with the intrinsic noise
background of the instrument, is to cross correlate the measurements $s_i$ of
multiple detectors.
When the background is assumed to be isotropic,
unpolarized and stationary, the cross correlation product is
given by \cite{all99}:

\begin{equation} 
Y=\int_{-\infty}^\infty \tilde{s_1}^*(f)\tilde{Q}(f)\tilde{s_2}(f) df
\end{equation}

where

\begin{equation}
\tilde{Q}(f)\propto \frac{\Gamma (f) \Omega_{\rm gw}(f)}{f^3P_1(f)P_2(f)}
\end{equation}

is a filter that maximizes the signal to noise ratio ($S/R$). In the
above equation, $P_1(f)$ and $P_2(f)$ are the power spectral noise
densities of the two detectors and $\Gamma$ is the non-normalized overlap
reduction function, characterizing the loss of sensitivity due to
the separation and the relative orientation of the detectors (see Fig.~\ref{fig-overlap}). 
The optimized $S/N$ ratio for an integration time $T$ is given by \cite{all99}:

\begin{equation}
(\frac{S}{N})^2 =\frac{9 H_0^4}{16 \pi^4}T  \int_0^\infty
df\frac{\Gamma^2(f)\Omega_{\rm gw}^2(f)}{f^6 P_1(f)P_2(f)}({\rm erfc}^{-1}(2
\beta)-{\rm erfc}^{-1}(2 \alpha))^{-2} .
\end{equation} 

\begin{table}
\centering 
\begin{tabular}{lccccc}
\noalign{\smallskip}
\hline
\noalign{\smallskip}
& LHO-LHO & LHO-LLO & LLO-VIRGO & VIRGO-GEO & EGO-EGO \\
\noalign{\smallskip}
\hline
\noalign{\smallskip}
initial& $2.8 \times 10^{-3}$ & $ 5.5 \times 10^{-6}$ &  $ 6.1 \times 10^{-6}$
&  $4.9 \times 10^{-6}$ & - \\
advanced & 0.52  & 0.029  & -  & -  & - \\
3$^rd$ generation & - & - & - & - & 8.4  \\
\noalign{\smallskip}
\hline
\end{tabular}
\caption{Expected signal-to-noise ratio, corresponding to the
  continuous background ($D>1$) and for the actual and future terrestrial
  interferometer pairs for an integration time
  T = 1 year, a detection rate $\alpha=90 \%$  and a false alarm rate
  $\beta=10\%$. LHO and LLO stand for LIGO Hanford Observatory and
  LIGO Livingston Observatory.}
\label{table-sensitivity}
\end{table}

\begin{figure}
\centering
\includegraphics[angle=0,width=0.8\columnwidth]{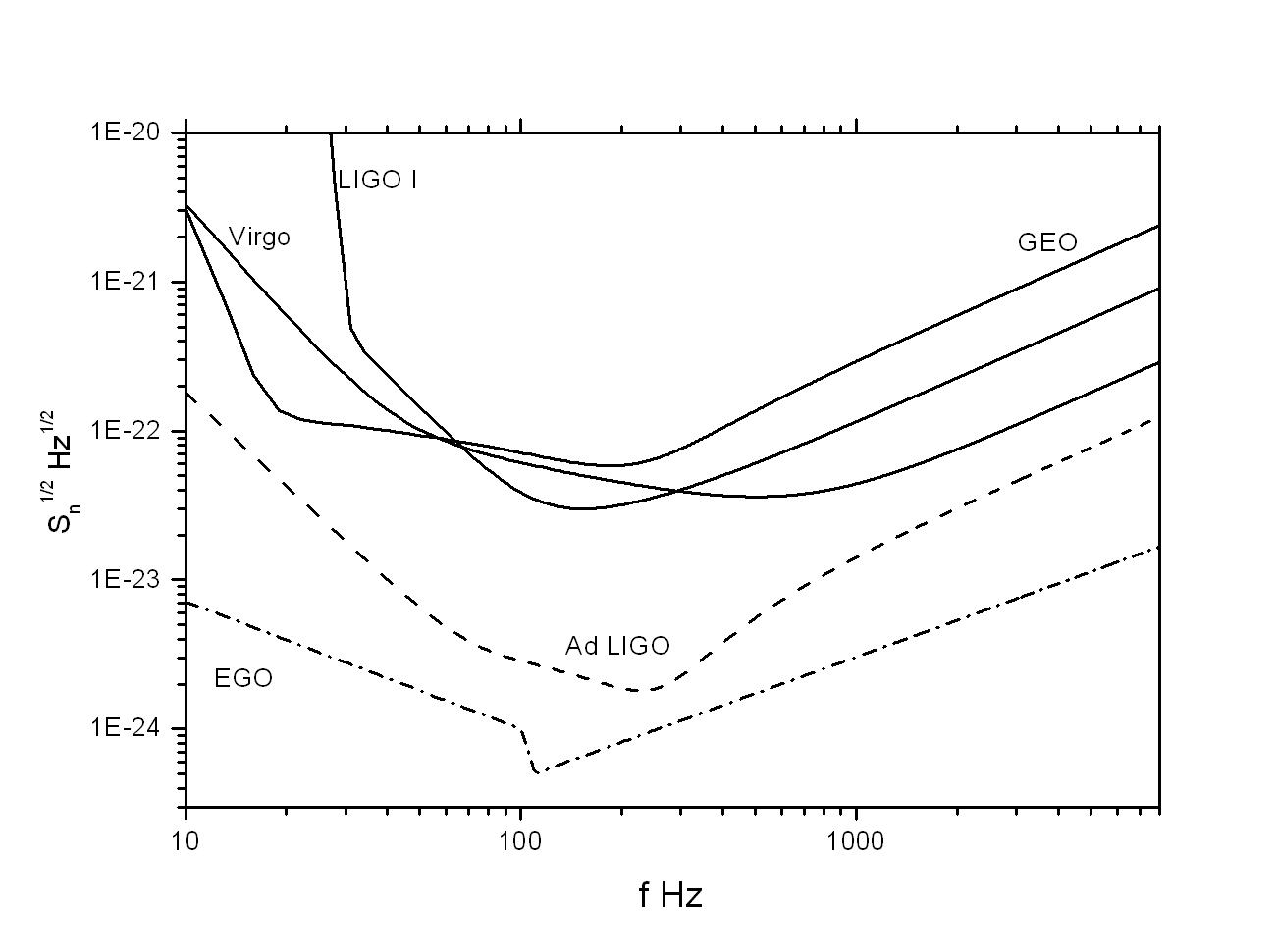}
\caption{designed sensitivities of the main first generation
  interferometers (continuous), compared to the planned sensitivities
  of the advanced interferometer LIGO Ad and the third generation
  interferometer EGO. 
\label{fig-noise}} 
\end{figure}

The $S/N$ for the  main
terrestrial interferometer pairs, at design sensitivity and in their
advanced configuration, after one
year of integration, are given in
Table~\ref{table-sensitivity}, for a detection rate $\alpha=90\%$ and
a false alarm rate $\beta=10\%$. Expressions for the power spectral
densities of actual detectors can be found in
\cite{dam01b} (see Fig.~\ref{fig-noise}).

The continuous signal is below the sensitivity that can be obtained by
cross-correlating actual pairs of detectors \cite{all99}.
For example, considering co-located and co-aligned interferometers,
such as Virgo or LIGO, we find a maximum signal-to-noise
ratio of $S/N \sim 0.003$ ($S/N \sim 0.5$) for the initial (advanced)
configuration. However, the sensitivity of the future third
generation of detectors such as EGO, presently in discussion, could be
high enough to gain one order of magnitude in the expected signal to
noise ratio ($S/N \sim 8$) . 
On the other hand, the popcorn noise contribution  
could be detected by new data analysis techniques currently under
investigation, such as the maximum likelihood statistic \cite{dra03}, 
or methods based on the Probability Event Horizon concept
\cite{cow05}, which describes the evolution, as a function of the
observation time, of the cumulated signal throughout the Universe.

\section{Simulations of the DNS population}

\begin{figure}
\centering
\includegraphics[width=1\columnwidth]{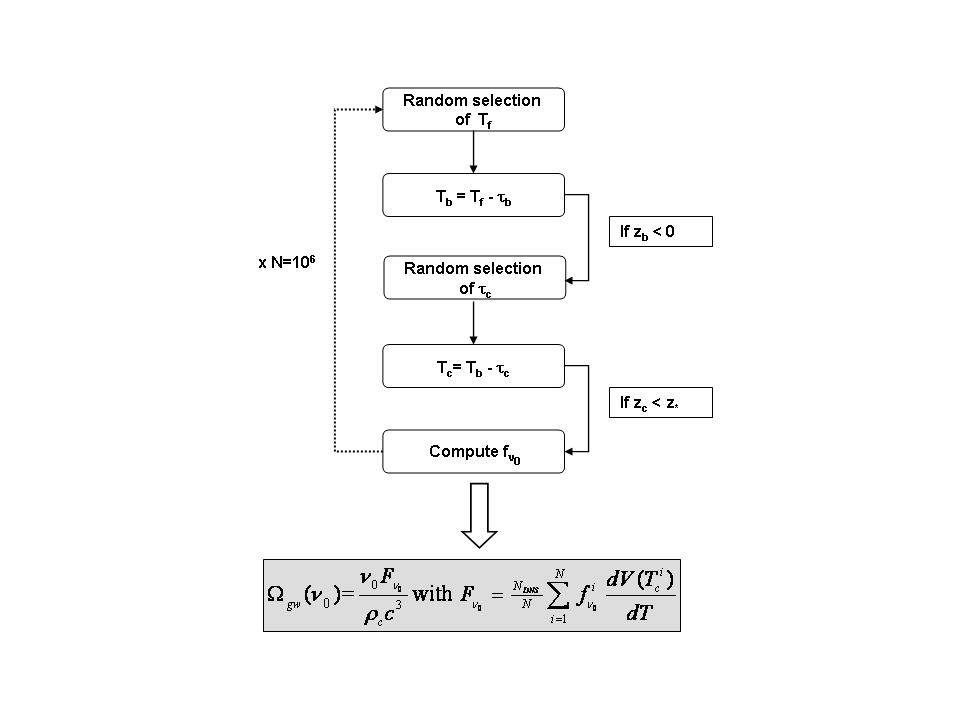}
\caption{flowchart of the Monte-Carlo simulations described in section
  3. Each system is generated with a cosmic time of formation $T_f$,
     (from which the cosmic time of formation of the DNS $T_b$ is calculated), 
     and a coalescence time $\tau_c$, which defines the coalecence cosmic time
     $T_c$. Only DNSs which coalesce at redshifts $z_c <
     z_*$ contribute to the integrated signal; their fluences are
     calculated and combined with adequate normalization factors (see
     text) to compute the total flux and the density parameter
     $\Omega_{gw}(\nu_o)$. The critical redshift to have a continuous stochastic background is $z_*=0.4$ ($z_*=0.2$ for the popcorn noise).  
\label{fig-flow}} 
\end{figure}

In this section, we introduce Monte Carlo simulations of the extra-galactic
population of DNSs and calculate the resulting stochastic
background. This method can be extended to any kind of sources, in
particular to GW events that are delayed with respect to the formation of the progenitors. 
The simulations follow the evolution of the
system from the birth of the progenitors to the merging of the two
neutron stars, after the redshift of
formation and the coalescence time have been selected. 
The difference with the previous simulations of \cite{reg06,reg07}, in
addition to the update of the initial mass function, the star
formation rate and the cosmological model, is that the normalization (the ratio between the real
number of DNSs and the number of simulated DNSs or runs) is done in a
more realistic way by considering comoving volume elements instead of
redshift intervals when following the evolution of the progenitors. 

To simulate a population of coalescences observed today in an element
of comoving volume, we proceed as follow (see Fig.~\ref{fig-flow}): 
\begin{enumerate} 
\item The time of formation of the progenitors is selected from
  the probability distribution
\begin{equation}
P_f(T_f)=\frac{R^o_{f}(T_f)} {\int_0^{T_{\max}}R^o_f(T_f) dT_f}
\label{eq-proba_frate}
\end{equation}
defined by normalizing in the interval $0-12.5$ Myr the formation rate
of the progenitors (eq.~\ref{eq-vrate_prog}), 

\item  The cosmic time $T_b$ at which the progenitors
have evolved into two neutron stars and start to coalesce is given by
\begin{equation}
T_b = T_f - \tau_b
\label{eq-Tb}
\end{equation}
where  $\tau_b$ ($\sim 10^8$ yr) has been defined in the previous
section as the mean lifetime of the progenitors.

\item The coalescence timescale which depends on both the orbital parameters
  and the masses of the two neutron stars, is selected from the
  probability distribution  eq.~\ref{eq-proba_tau}, between 0.2 Myr and
  20 Gyr. 

\item The cosmic time $T_c$ at which the coalescence occurs is given by
\begin{equation}
T_c = T_b - \tau_c
\label{eq-Tc}
\end{equation}
and the corresponding coalescence redshift  $z_c$ is derived by solving the equation:
\begin{equation}
T_c = \int^{z_c}_{0} \frac{dz}{H_0(1+z)E(z)}
\label{eq-zc}
\end{equation} 
 
\item Each DNS, thus generated, is then sorted into bins of cosmic time
$[T_c^j;T_c^j+\Delta T_c]$, for which we calculate the total flux as the sum
of all the individual fluences, normalized by the ratio between the
total formation rate of the progenitors in the range $0-12.5$ Gyr
($N_p^o$) and the number of simulated DNSs ($N_{sim}$):
\begin{equation}
F_j(T^j_c,\nu_o) = \frac{N_p^o}{N_{\rm{sim}}} \sum_{i} f(T^i_c,\nu_o)
\end{equation}
with $T^i_c$ in the range $[T_c^j;T_c^j+\Delta T_c]$ 
and where
\begin{equation}
N_p^o= \int_0^{T_{\max}} R^o_f(T_f) dT_f
\end{equation}

\item The model of the star formation rate being isotropic, the element
  of comoving volume at cosmic time $T$ is considered as representative
  of the entire population in the shell between $[T;T+dT]$.
Therefore, the total flux from sources located between $[T_c^j;T_c^j+\Delta T_c]$ writes:
\begin{equation}
F(\nu_o) = \sum_j F_j(T^j_c,\nu_o) \frac{dV}{dT}(T^j_c)
\end{equation}
or equivalently
\begin{equation}
F(\nu_o) = \frac{N_p^o}{N_{\rm{sim}}} \sum_{i} f(T^i_c,\nu_o)\frac{dV}{dT}(T^i_c)
\end{equation}
%It results for the closure density
%\begin{equation}
%\Omega_{gw}= 8.6 \times 10^{-10} \nu_o^{2/3}  \frac{N_p^0}{N_{\rm{sim}}}
%\sum_{i} (1+z_c^i)^{-1/3}
%\label{eq-omega_DNS}
%\end{equation} 
\end{enumerate}

\begin{figure}
\centering
\includegraphics[angle=0,width=0.8\columnwidth]{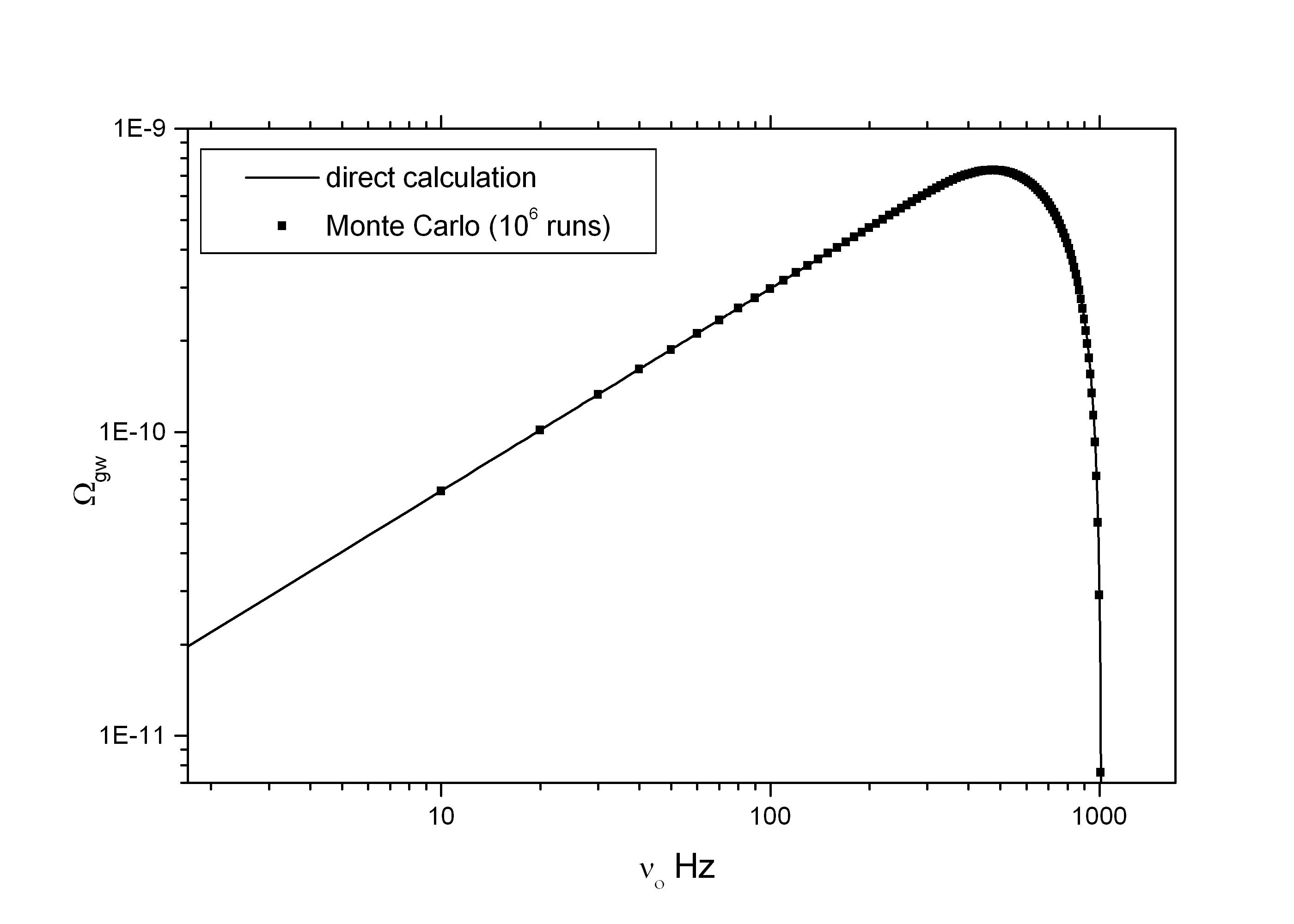}
\caption{closure density of the continuous background produced by DNS
  coalescences at $z > 0.5$ derived from the direct calculations described
  in section 1 (continuous line) and from the Monte Carlo simulations
  described in section 2 (square), for a number of runs
  $N_{sim}=10^6$. The agreement between the two is better than $99.5\%$.
\label{fig-omega_MC}} 
\end{figure}

The closure density (eq.~\ref{eq-omega_astro}) corresponding to the
continuous background ($z_c>0.5$) is plotted in figure
(Fig.~\ref{fig-omega_MC}) and compared with the results obtained in
section 1.
For a number of runs $N_{sim}=10^6$, the agreement is better than
$99.5\%$, which is accurate enough to validate the Monte Carlo
procedure.

\section{Conclusions}
We presented Monte Carlo numerical simulations of the extra-galactic
population of double neutron stars and investigated its contribution
to the gravitational wave stochastic background.
The stochastic background formed by the final stage of the coalescence
in the frequency band $10-1500$ Hz is continuous for sources beyond
$z \sim 0.5$ and rather a popcorn noise between $0.2<z<0.5$.
The closure density of the continuous contribution reaches a maximum of
$\Omega_{gw} \sim 7.3 \times  10^{-10}$  at around  480 Hz, which is
below the sensitivity of actual and advanced interferometers
but may be detectable with the third generation.
The popcorn contribution seems more promising with $\Omega_{gw} \sim 9.7
\times  10^{-10}$ at 520 Hz.
The advantage of the Monte Carlo simulations, compared to the direct
calculation are numerous. On the one hand they permit to study the statistical
properties of the background in the time domain, in particular the non
gaussian popcorn contribution, for which adapted detection strategies
are currently under development. The
simulation of GW time series \cite{cow02a,cow05b} that can be injected
in the output of a pair of detectors \cite{bose03,cel07}, is essential
to test and validate data analysis pipelines. 
On the other hand incorporating new parameters (such as the
eccentricity, the orbital separation and the masses of the two stars)
can be done in a very simple way. In this work all the initial
informations are included in the coalescence time. 
However to investigate the stochastic background formed by the low
frequency inspiral phase, in the frequency domain of the spatial
detector LISA, when the system can be highly eccentric \cite{cha05,ihm06}
emitting GW at higher harmonics to the keplerian frequency
\cite{pet63,pet64,ign01}, one needs to follow the combined evolution
of the frequency, the eccentricity and the redshift of emission.
This work is currently in progress and will be reported in a future paper.

\end{document}